\documentclass[aps,prx,longbibliography,superscriptaddress,twocolumn,showkeys]{revtex4-2}

\DeclareUnicodeCharacter{2032}{\ensuremath{^{\prime}}}

\usepackage{graphicx, color}
\usepackage{xcolor}
\usepackage{amsfonts,amssymb,amsmath}
\usepackage{siunitx}
\usepackage{bm}
\usepackage{bbold}
\usepackage{braket}
\usepackage{mathtools}
\usepackage{dsfont}
\usepackage{booktabs}

\usepackage[colorlinks=true, linkcolor=blue, citecolor=blue, urlcolor=blue]{hyperref}

\begin{document}

\title{Topological insulator single-electron transistors for charge sensing applications}

\author{Omargeldi Atanov}
\thanks{These authors contributed equally to this work. \newline Correspondence: \href{mailto:atanov@ph2.uni-koeln.de}{atanov@ph2.uni-koeln.de}}
\affiliation{Physics Institute II, University of Cologne, D-50937 K{\"o}ln, Germany}

\author{Junya Feng}
\thanks{These authors contributed equally to this work. \newline Correspondence: \href{mailto:atanov@ph2.uni-koeln.de}{atanov@ph2.uni-koeln.de}}
\affiliation{Physics Institute II, University of Cologne, D-50937 K{\"o}ln, Germany}

\author{Jens Brede}
\affiliation{Physics Institute II, University of Cologne, D-50937 K{\"o}ln, Germany}

\author{Oliver Breunig}
\affiliation{Physics Institute II, University of Cologne, D-50937 K{\"o}ln, Germany}

\author{Yoichi Ando}
\affiliation{Physics Institute II, University of Cologne, D-50937 K{\"o}ln, Germany}

\begin{abstract}
We present topological insulator (TI)-based single-electron transistors (SETs) as magnetic-field-compatible charge sensing devices that are easily integrable with TI-superconductor hybrid platforms. We observe well-resolved Coulomb diamonds in the charge-stability diagrams of our devices confirming the charge quantization and single-electron transport. In some devices, the Coulomb resonances show persistent shifts corresponding up to $\sim$ $e/2$ charge. An axial magnetic field further displaces these shifts to higher or lower gate voltages. We find that the axial magnetic-field dependence of the shifts is consistent with the Zeeman shift of a trap state coupled to the SET, and we reproduce the observations using numerical simulations. The resonance shifts are therefore identified as a consequence of the sensitivity of our TI-SET devices to charges in proximity. Establishing this charge sensing capability is a first step toward integrating TI-SETs as charge sensors in more complex TI-based hybrid devices, with the overarching goal of detecting and braiding Majorana zero modes.
\end{abstract}

\maketitle

\newpage

\section{Introduction}

Electronic transport through a conductive island with sufficient charging energy is highly influenced by the surrounding electrostatic environment. Due to this sensitivity, single-electron transistors (SETs) and quantum dots (QD) have emerged as excellent charge sensors both in DC \cite{dresselhaus1994measurement, keller1996accuracy} and radio-frequency (RF) \cite{schoelkopf1998radio, aassime2001radio} regimes of transport. In nanoscale semiconductor devices, it is a common practice to fabricate a QD charge sensor from the same material system as the device under investigation \cite{hu2007ge, mannila2019detecting, razmadze2019radio, vandriel2024charge, van2026single}. This gives the advantage of controlling the device--sensor coupling via gate voltages and reduces the number of fabrication steps. In this context, fabricating an SET charge sensor based on a TI material opens up an exciting route toward integration with TI-superconductor hybrid devices as a non-invasive probe of topological phase transitions \cite{hegde2020topological, ben2015detecting, razmadze2019radio}. For instance, a TI-based SET could be readily integrated with the RF-SQUID type TI-Al junctions \cite{schluck2024robust} or the TI nanowire (TINW)-based columnar nano-SQUID devices \cite{nikodem2024topological}, both of which are promising device architectures for exploring topological superconductivity. Correlated signatures from tunneling spectroscopy and charge sensing could help identify topological phase transitions accurately.

Bulk-insulating TI materials have gapless and non-degenerate spin-momentum locked surface states that are robust against non-magnetic backscattering due to time-reversal symmetry. Owing to the Dirac nature of the electrons on the surface of a TI, a purely gate-defined confinement is prevented by Klein tunneling \cite{katsnelson2006chiral}. An alternative method of confinement is via physical constrictions that hybridize the surfaces of the TI and open a gap at the Dirac point \cite{zhang2010crossover, munning2021quantum}. In this context, a TI-based quantum dot was demonstrated in Bi$_2$Se$_3$ by using ultrathin parts of the sample as tunnel barriers \cite{cho2012topological}, although large back-gate voltages were still required to observe Coulomb blockade. A later work realized a Bi$_2$Te$_3$-based SET using focused ion beam--etched constrictions \cite{jing2019single}. Despite high degree of control in fabrication, this approach is not readily integrable as a charge sensor with other device architectures. Furthermore, the stability of Coulomb resonances under an applied magnetic field has not been studied in either platform. This is particularly relevant for the integration of SET charge sensors with TI–superconductor hybrid devices that require magnetic fields to reach topological regimes \cite{cook2011majorana, schluck2024robust, nikodem2024topological}.

In this work, we report highly reproducible SET devices based on bulk-insulating TI material BiSbTeSe$_2$ (BSTS2) \cite{arakane2012tunable} and AlO$_x$ tunnel barriers. We demonstrate the charge sensing capability as well as the magnetic-field compatibility of our devices by detecting the non-stochastic charging events of a localized charge trap at axial magnetic fields up to 6 T. These results establish TI nanowire SETs as sensitive charge detectors and demonstrate their suitability for integration with TI-based hybrid quantum devices.


\begin{figure*}
    \centering
    \includegraphics[width=\textwidth, trim={0cm 0cm 0cm 0cm},clip]{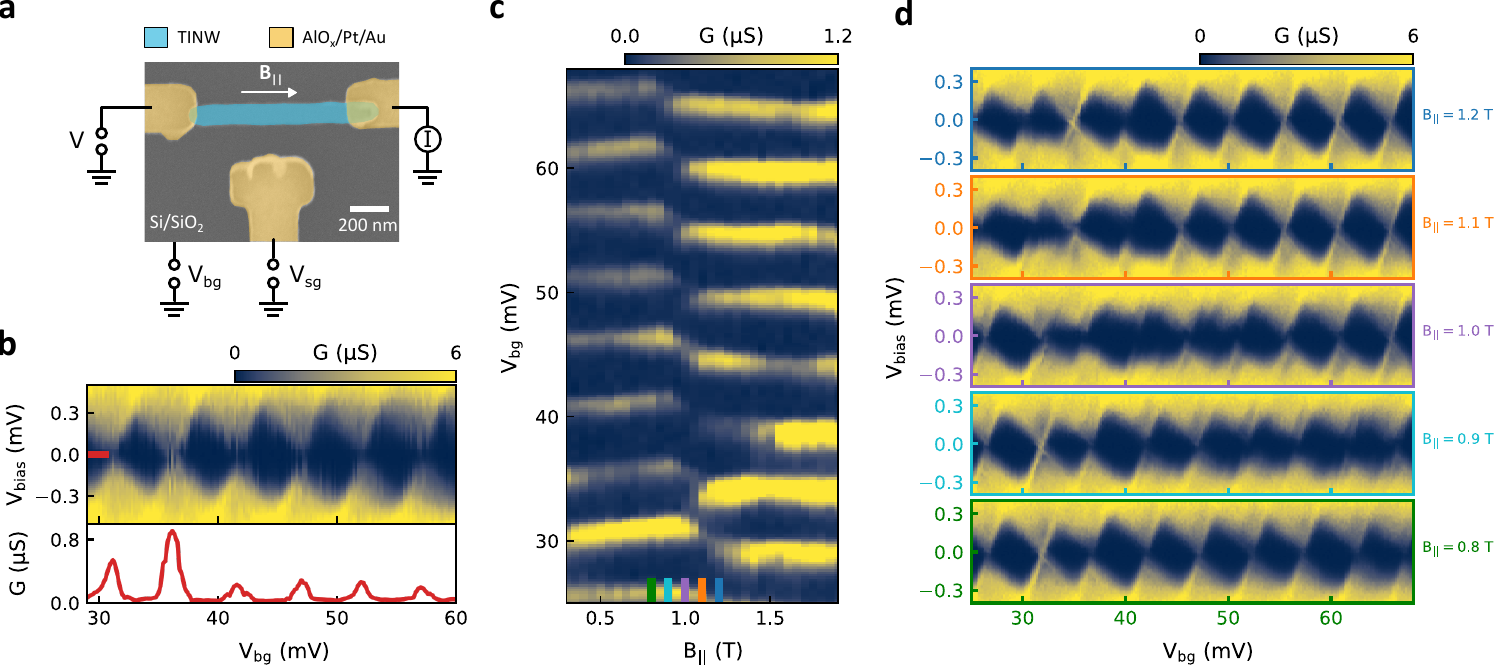}
    \caption{Topological insulator SET and its charge-stability diagram. \textbf{a} False-colour scanning electron microscope (SEM) image of the SET device made of BSTS2 nanowire (sky blue) contacted by Pt/Au electrodes (yellow) at two ends.
    \textbf{b} Typical charge-stability diagram of the TI-SET showing differential conductance $G$ as a function of $V_{\mathrm{bias}}$ and $V_{\mathrm{bg}}$. It exhibits similar sized Coulomb diamonds across several charge degeneracy points tuned by the back-gate voltage. The bottom panel shows a zero-bias gate trace of the conductance taken from the upper panel. \textbf{c} Color plot of zero-bias conductance as a function of $V_{\mathrm{bg}}$ and $B_{||}$ showing the evolution of Coulomb resonances with axial magnetic field $B_{||}$. At around 1 T, the resonances shift to lower $V_{\mathrm{bg}}$ values while preserving the peak-to-peak spacing.
    \textbf{d} Charge-stability diagrams of of the SET at various $B_{||}$ values close to 1 T. The diamonds are distorted at the $V_{\mathrm{bg}}$ values where the zero-bias resonances show shifts in panel \textbf{c}.
    }
    \label{fig:Fig1}
\end{figure*}

\section{Results}

\subsection{Topological Insulator Single-Electron Transistor}

Fig. \ref{fig:Fig1}\textbf{a} shows the false-colored SEM image of a typical TI-SET device along with the measurement configuration. To fabricate the SETs, BSTS2 flakes are deterministically transferred to a doped Si substrate covered with 300 nm SiO$_2$ layer. The flakes are etched into TINW islands with widths 100-150 nm and lengths 1-2 $\mu$m as described in Ref. \cite{feng2025long}. The leads and the side-gate electrode are patterned by electron beam lithography, after which the exposed regions are covered with a few layers of AlO$_x$ using an atomic layer deposition system. Metallic electrodes (5~nm Pt / 35~nm Au) are then sputtered. The chemical potential of the TINW island is modulated by applying a back-gate voltage $V_{\mathrm{bg}}$ to the Si substrate. Applying voltage $V_{\mathrm{sg}}$ to the side-gate allows the same modulation with a different lever arm. Except the data in Fig. S4 of the Supplementary Information, all measurements in this work are performed using only the back-gate. We report results from a single device (Device 1) throughout the paper and we observe clear Coulomb-blockaded transport in 3 more devices fabricated identically (see Supplementary Fig. S2).

\begin{figure*}
    \centering
    \includegraphics[width=\textwidth, trim={0cm 0cm 0cm 0cm}, clip]{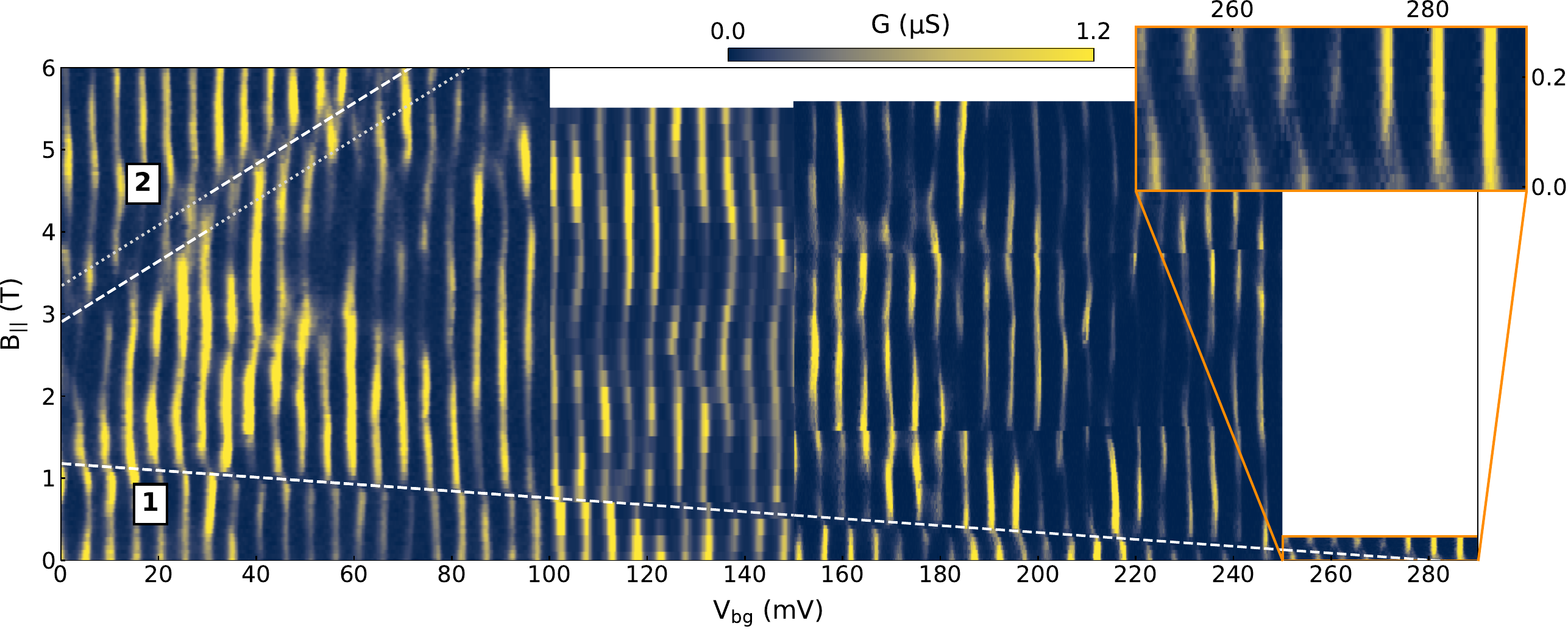}
    \caption{Magnetic-field dependence of the Coulomb resonances in a large $V_{\mathrm{bg}}$ range.
    The energy of the trap state shifts linearly with in-plane magnetic field. In the first region, denoted by `1', the slope of the linear line is negative revealing a spin orientation in the opposite direction of the $B_{||}$. In the second region, denoted by `2', the two parallel lines with positive slope correspond to the same trap state which has a spin polarization in the direction of $B_{||}$. The trajectory followed by the trap signature is shown with a dashed line. The inset shows the fine shifts of the resonances in region `1' that correspond to $\sim$ $e/5$ charge at low $B_{||}$ field.}
    \label{fig:Fig2}
\end{figure*}

Charge-stability diagram of the TI-SET, shown in Fig. \ref{fig:Fig1}\textbf{b}, exhibits well-defined Coulomb diamonds. Similar sized diamonds suggest that there is a single dominant charging energy and that the regime of transport in this device is the classical Coulomb blockade. A zero-bias linecut plotted in the bottom panel of Fig. \ref{fig:Fig1}\textbf{b} shows alternating Coulomb resonance peaks separated by regions of vanishing conductance.

From the size of the Coulomb diamonds we can estimate the charging energy $E_C$ of the island. The addition energy of the island $E_{add} = 2E_{C} + E_{N}$ has two contributions, one is the classical charging energy $E_{c}$ and the other is the single particle level spacing $E_{N}$ resulting from quantum confinement. As evidenced by the FFT analysis of the conductance oscillations (see Supplementary Fig. S1), the spacing between the Coulomb peaks does not show significant variation suggesting that the $E_C$ is the only relevant energy. Therefore, we can assume $E_{add} \sim 2E_{C}\sim 0.33$ meV. The total capacitance is then $C_{\Sigma} = C_{s} + C_{d} + C_{g} = e^2/2E_{C} \approx 243$~aF. The geometric back-gate capacitance can be calculated analytically following Ref. \cite{wunnicke2006gate}. For a nanowire of length $L=900$~nm and width $W=120$~nm with a rectangular cross-section, we obtain $C_{g, geo}$ = 26.3 aF. The back-gate lever arm of the island is given by $\alpha=2E_C/e\delta V_{\mathrm{bg}}=C_{g}/C_{\Sigma}\approx 0.13$ which yields $C_{g} \approx 32$ aF considering $\delta V_{\mathrm{bg}} \approx 5$ mV as extracted from the FFT analysis. The small discrepancy between the measured gate capacitance and the calculated geometric capacitance could be due to the non-uniform shape of the nanowire.

\subsection{Coulomb resonance shifts}

Fig. \ref{fig:Fig1}\textbf{c} shows the magnetic-field dependence of the Coulomb resonances in a small $V_{\mathrm{bg}}$ range. $B_{||}$ is applied along the wire direction, as indicated in Fig. \ref{fig:Fig1}\textbf{a}. Clear shifts of the resonances can be observed at magnetic fields close to 1 T. To investigate these shifts in more detail, we performed Coulomb blockade spectroscopy of the SET at several B$_{||}$ values around 1 T as shown in Fig. \ref{fig:Fig1}\textbf{d}. Beyond the usual Coulomb diamonds, these plots reveal distortions in some diamonds at $V_{\mathrm{bg}}$ values corresponding to the resonance shifts. The position of the distortion signature in $V_{\mathrm{bg}}$ axis is furthermore changing with the magnetic field. It first occurs around 60 mV in the 0.9 T panel and it continues to shift down to 30 mV in the 1.2 T panel. This type of distortion in the stability diagram is commonly caused by a nearby stray charge that influences the electrostatic potential of the SET island. On the other hand, considering the fact that the data in Fig. \ref{fig:Fig1}\textbf{c} and \textbf{d} takes several hours to acquire, the stability of the signature is remarkable suggesting the mechanism behind it is non-stochastic. Out of the 4 devices measured, we found the same resonance shifts in one more device (see Supplementary Fig. S3).

\begin{figure}
    \centering
    \includegraphics[width=0.48\textwidth, trim={0cm 0cm 0cm 0cm},clip]{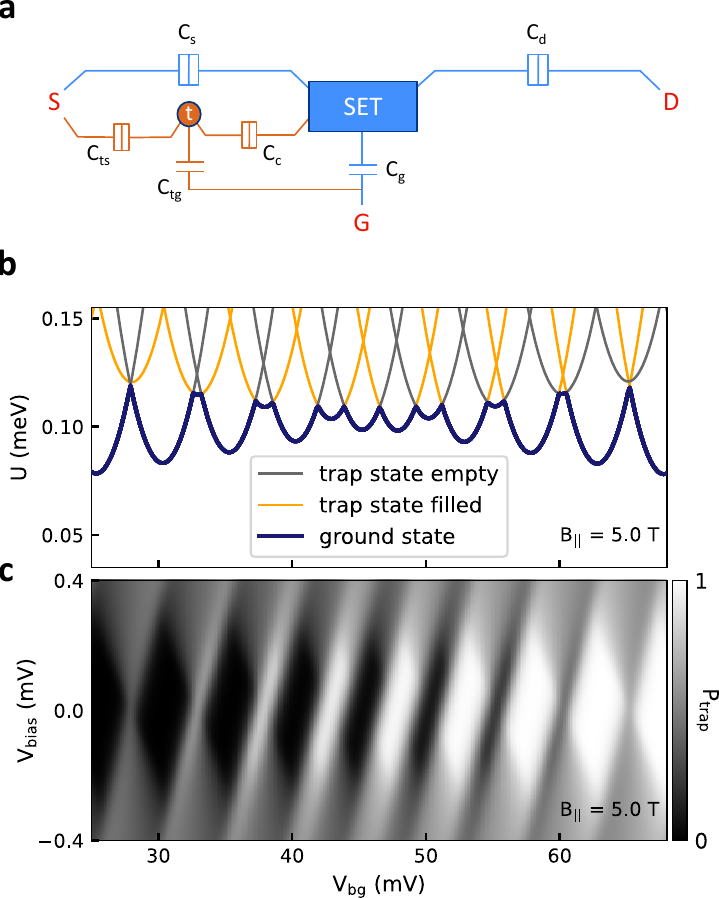}
    \caption{\textbf{a} Equivalent circuit diagram of the main SET island tunnel-coupled to a charge trap that mimics a small quantum dot. \textbf{b} Total electrostatic energy of the system follows one of the two sets of charge parabolas depending on whether the relevant trap state is empty (gray line) or occupied (orange line). The ground state switches from empty to occupied configuration near the trap state degeneracy. \textbf{c} The mean occupation probability of the trap state as a function of $V_{\mathrm{bias}}$ and $V_{\mathrm{bg}}$. The probability fluctuates near the trap state degeneracy corresponding to the switching of the ground state of the total system.
    }
    \label{fig:Fig3}
\end{figure}

\begin{figure*}
    \centering
    \includegraphics[width=\textwidth, trim={0cm 0cm 0cm 0cm}, clip]{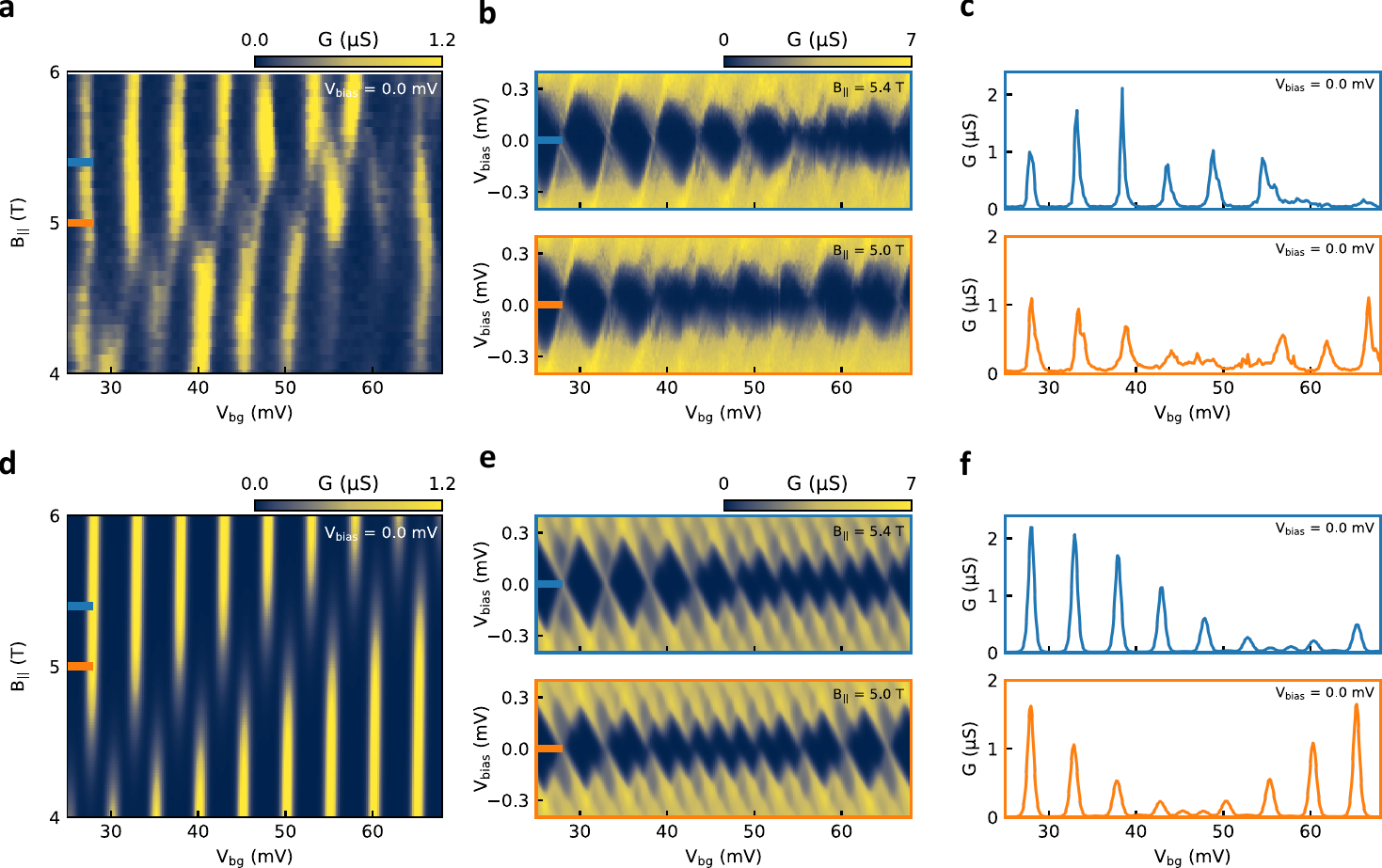}
    \caption{Comparison between the measured and simulated conductance data. \textbf{a} Zero-bias Coulomb resonances as a function of $B_{||}$ in the region `2' shown in Fig. \ref{fig:Fig2}. The resonances shift to higher $V_{\mathrm{bg}}$ values as the $B_{||}$ is increased. \textbf{b} Charge-stability diagram of the TI-SET at $B_{||}=5.0$ T (lower panel) and at $B_{||}=5.4$ T (upper panel). The diamond distortions are again visible in the vicinity of the resonance shifts observed in \textbf{a}. \textbf{c} Linecuts of the two panels in \textbf{b} taken at zero-bias voltage. In each of the two panels the suppression of the conductance peaks occurs at the $V_{\mathrm{bg}}$ values where the distorted diamonds are observed in their corresponding panels in \textbf{b}. \textbf{d} Simulated conductance plot of a SET-trap system that includes the Zeeman shift of the trap energy. \textbf{e} Numerical simulation of the stability diagram of the SET-trap model at two different magnetic fields; $5.0$ T in the lower panel and $5.4$ T in the upper panel. The trap degeneracy is shifted by the magnetic field due to Zeeman effect. \textbf{f} Linecuts taken from the two panels in \textbf{e} at zero-bias voltage indicated by thick orange and blue ticks. See Supplementary Information for the list of parameter values used in the simulation.}
    \label{fig:Fig4}
\end{figure*}

Fig. \ref{fig:Fig2} shows about 57 Coulomb resonances in a large $V_{\mathrm{bg}}$ and B$_{||}$ field range. Note that this figure was produced by combining measurements taken on different days and with different number of data points. Therefore, the color contrast and the resolution vary slightly across the figure. The resonance shifts shown in Fig. \ref{fig:Fig1}\textbf{c} are now visible in an extended gate-voltage range in the region `1'. These shifts can be traced down to zero magnetic field using a linear line (see inset for a low-field zoom-in). The linear evolution of the signature position in energy with $B_{||}$ field points to the role of the Zeeman effect. In fact, similar signatures observed in a silicon nanowire were attributed to the Zeeman shift of a singly occupied trap state that is tunnel- coupled to the nanowire \cite{hofheinz2006individual}. Resonance shifts due to charge traps and localized states in various other platforms were also reported \cite{grupp2001dynamical, guttinger2009electron, zimmerman1997modulation}. In light of this, we also attribute the observed resonance shifts to the charging event of a trap state coupled to the SET. Since we can systematically trace the shifts in $V_{\mathrm{bg}}$ axis, it is safe to assume that the charging of the trap is controlled by the gate-voltage and is deterministic \cite{grupp2001dynamical}. Applying magnetic field shifts the energy of the trap state due to Zeeman effect and results in the shift of the trap signature along the $V_{\mathrm{bg}}$ axis.

Another set of resonance shifts is observed in region `2'. In this region, the resonances move toward higher gate-voltage with increasing $B_{||}$, indicating a trap state with spin orientation opposite to that in region `1'. Notably, the shifts follow two trajectories: at lower fields they trace the lower line, before switching to the upper line at approximately 4~T. The two lines have the same slope and the shifts are continuous in $V_{\mathrm{bg}}$ axis, suggesting that both trajectories originate from the same trap. The transition from the lower line to the upper one indicates a modification of the effective gate lever arm around 4~T, likely due to changes in the local electrostatic environment.

Zeeman shift of the trap signature with magnetic field can be fit using the following expression \cite{hofheinz2006individual}:

\begin{equation}
    e\alpha_t\frac{\partial V_{\mathrm{bg}}}{\partial B_{||}} = g\mu_B\Delta S_\mathrm{z}
\end{equation}

\noindent where $\Delta S_\mathrm{z}$ denotes the change in spin state of the occupied level in the trap and $\alpha_t$ is the lever arm of the trap characterizing its coupling to the gate electrode. The slopes of the linear lines shown in Fig. \ref{fig:Fig2} can therefore be used to extract the gate lever-arm of the traps and the spin orientation of the electron added to the trap. The estimated slope of trap `1' is $({\partial B_{||}/\partial V_{\mathrm{bg}}})_{1} = -4.2$ $\mathrm{T/V}$ resulting in $\alpha_t = 1.38 \times10^{-5}$. Similarly, the lever arm for the trap `2' can be calculated using its slope $({\partial B_{||}/\partial V_{\mathrm{bg}}})_{2} = 37.14$ $\mathrm{T/V}$ giving $\alpha_t = 2.15 \times10^{-3}$. In both cases we have assumed $\textit{g} = 2$ and single occupation of the trap state leading to the spin quantum number to change by $\Delta S_\mathrm{z}=\pm1/2$.

Because the resonance shifts originate from the charge induced on the SET island by an additional electron in the trap, the capacitive coupling between the TINW and the trap can be determined from the relation:

\begin{equation}
    \Delta V_{\mathrm{bg}} = \beta_t \delta V_{\mathrm{bg}}
\end{equation}

\noindent where $\Delta V_{\mathrm{bg}}$ is the shift of the resonance in gate voltage and $\beta_t$ is the lever arm characterizing the coupling between the TINW island and the trap. Using this relation, we can estimate $\beta_t$ which is a useful parameter to identify the position of the trap. For the trap `2' in Fig. \ref{fig:Fig2}, the resonance shifts are as high as $\Delta V_{\mathrm{bg}} \approx 2.5$ mV resulting in $\beta_t = 0.5$. Similarly, for the trap `1' with opposite spin state, the $\Delta V_{\mathrm{bg}} \approx 2.3$ mV which leads to $\beta_t = 0.46$.

\subsection{Numerical simulation of a trap-coupled SET}

To support our interpretation of the observed signatures, we numerically simulate electronic transport through a trap-coupled SET \cite{hofheinz2006individual, villis2014direct}. The equivalent circuit is shown in Fig.~\ref{fig:Fig3}\textbf{a}. We consider a metallic SET island with negligible single-particle level spacing that is tunnel-coupled to a localized trap effectively behaving as a small quantum dot. The trap is tunnel-coupled to the source electrode and capacitively coupled to the gate electrode. The electrostatic energy of the coupled system can be written as:
\begin{equation}
    U(n, n_t)=E_C(N-n+\beta_t(N_t-n_t))^2 + E_t(N_t-n_t)^2
\end{equation}
where $E_C$ is the charging energy of the SET island and $E_t={e^2}/[2(C_{tg}+C_{ts}+C_c)]$ is the charging energy of the trap. Here, $N$ and $N_t$ denote the induced charges on the SET island and the trap, respectively.

Using this energy expression to evaluate the tunneling rates, we solve the Pauli master equation self-consistently to obtain the stationary probabilities of all relevant charge states labeled by $(n,n_t)$, where $n\in\{0,\dots,9\}$ is the integer charge on the SET island and $n_t\in\{0,1\}$ denotes the occupation of the trap charge state whose degeneracy lies within the simulated gate-voltage range. Near a given charge degeneracy point, the dynamics are restricted to a small set of charge transitions, which allows us to describe the system in terms of relative occupation numbers rather than the total charge on the SET island and the trap. In this framework, the trap is modeled by tracking the occupation probability of a single addition transition. The mean source–drain current is obtained by summing over all allowed transitions between the accessible charge states within the sequential tunneling model \cite{beenakker1991theory}. Further details of the simulation are provided in the Supplementary Information.

By employing this model, we first calculate the gate-voltage dependent energy of the SET coupled to trap `2' at $B_{||}=5.0$~T to demonstrate the ground-state transition of the coupled system due to the charging of the trap state. We obtain two distinct charge parabolas corresponding to empty and occupied trap state as shown in Fig. \ref{fig:Fig3}\textbf{b}. The system follows the lowest energy transitions that occur at the crossing points between neighboring parabolas. The ground state of the system changes from empty to occupied trap state configuration in a gradual manner when the gate-voltage is swept across the trap state degeneracy of $\sim$~46.5 meV. During this transition, the mean occupation of the trap state fluctuates between 0 and 1 electron as shown in Fig. \ref{fig:Fig3}\textbf{c}. At low-bias voltages, the crossings between gray and orange parabolas do not give rise to a current through the system. This is because the source-drain conduction is only possible when the degeneracy points of the gray (orange) parabolas are reached. The ground state of the system, however, traces lower energy crossings in this region. Current through the SET is restored once the ground state energy reaches the degeneracy points of the orange parabolas.

In Fig. \ref{fig:Fig4}, we present a detailed investigation of the resonance shifts at region `2' by comparing the measured data (\textbf{a}-\textbf{c}) with numerical simulations (\textbf{d}-\textbf{f}). Fig. \ref{fig:Fig4}\textbf{a} and \textbf{d} show the evolution of the zero-bias resonances as the field is increased from 4 T to 6 T. The simulated conductance of our SET-trap model accurately reproduces the resonance shifts observed in the measurements. Here we used the experimental values of $\alpha_t$ and $\beta_t$ but shifted the trap energy by a constant amount such that the signatures would appear in the same $(V_\mathrm{bg},B_{||})$ window as the experiment. It is important to note that our model could not reproduce the non-zero conductance in the trap degeneracy region that produces a splitting resonance peak in the data. This feature is only present in region `2' and we have not observed such a splitting in region `1' as evidenced by \ref{fig:Fig1}\textbf{c}. We believe that this effect can only be captured by considering higher order tunneling processes such as elastic co-tunneling. Furthermore, there could be a finite coupling between the trap and the drain lead, as a result of which a finite current can flow through the trap.

Fig. \ref{fig:Fig4}\textbf{b} and \textbf{e} show the measured and simulated stability diagrams, respectively, at two different $B_{||}$ fields. Due to the Zeeman effect, the degeneracy of the trap shifts to higher energy. As a result, the trap becomes occupied at a higher gate-voltage which in turn shifts the ground state transition to a higher gate-voltage. This explains the shifts of the distorted regions in Fig. \ref{fig:Fig4}\textbf{b}, and therefore confirms that our SET-trap model captures the mechanism of the observed distortions. The individual linecuts taken at zero-bias voltage of the stability diagrams are shown in Fig. \ref{fig:Fig4}\textbf{c} and \textbf{f}. The suppression of the conductance peaks in the ground-state transition regions can be clearly seen in both experimental and numerical data.

\section{Discussion}

The exact spatial location and microscopic origin of the charge traps cannot be determined unambiguously in our setup. Nevertheless, we can infer their likely location by comparing various capacitances we have extracted. Significantly larger values of $\beta_t$ compared to $\alpha_t$ indicate that the capacitive coupling between the trap and the TINW island is much stronger than the coupling to the gate electrode. This is also directly evident from the capacitance values used in the simulations ($C_{tg} = 0.43$~aF and $C_c = 100$~aF) which are chosen to yield quantitative agreement with the experimental data. Given that the dielectric constant of the TINW ($\epsilon \approx 200$)~\cite{borgwardt2016self} is substantially larger than that of SiO$_2$ ($\epsilon = 3.9$) and AlO$_x$ ($\epsilon = 9$), the trap is more likely located within the nanowire rather than in the gate dielectric or tunnel barriers.

A natural candidate for such traps is charge puddles arising from compensation doping in BSTS2~\cite{knispel2017charge, brede2024characterizing, borgwardt2016self}. These puddles form locally $p$- or $n$-doped regions due to disorder-induced potential fluctuations and band bending. Although charge puddles are less likely to form within a very limited spatial extent, such as in a short nanowire, they can appear in regions where the local potential variation is sufficiently strong. If such a localized puddle is created in a TI nanowire, it may host quantized charge. Scanning tunneling microscopy studies revealed a characteristic size of 40--50~nm for the surface puddles in BSTS2 samples \cite{knispel2017charge}. Within a simple parallel-plate capacitor model for the trap--gate coupling, the capacitance value used in the simulations provides an estimate of the trap size. Assuming a two-dimensional square geometry, a capacitance $C_{tg} = 0.43$~aF corresponds to a side length $l \approx 60$~nm which is consistent with reported surface charge puddle sizes in BSTS2.

The observation of both spin-up and spin-down single particle states, as in Fig. \ref{fig:Fig2}, demonstrates that the charge puddles can accommodate multiple electrons. However, the very small gate lever arms suggest that the measured gate-voltage range corresponds to only a small fraction of the trap charging energy. For instance, the full 290\,mV gate-voltage sweep translates to an electrochemical potential change of approximately 4\,$\mu$eV for trap `1', nearly two orders of magnitude smaller than a typical charging energy expected for a nanometer-scale trap. As a result, only a limited portion of the addition spectrum is accessed in our measurements, and the complete occupation dynamics of the puddles cannot be resolved.

We emphasize that the rare formation of charge puddles exhibiting non-stochastic charging does not hinder the use of our TI-SET devices as integrated charge sensors. Our data show that, in the $(V_\mathrm{bg}, B_{||})$ plane, one can always find regions that are free of trap dynamics. Furthermore, tuning of the TINW island potential via a local side-gate enables an independent calibration of the SET prior to probing the charge dynamics of a coupled device. One can therefore select a gate-voltage range corresponding to stable Coulomb resonances and operate the SET reliably as a charge sensor at a given magnetic field.

In summary, we demonstrate TI-SETs as magnetic-field compatible charge sensors for TI-superconductor hybrid platforms. The devices exhibit metallic SET behavior with stable Coulomb resonances in axial magnetic fields as high as 6~T. Supported by numerical simulations, we further resolve non-stochastic charging events of a trap state tunnel-coupled to the SET. These results pave the way for integrating TI-based SET charge sensors with TI Majorana platforms.

\section{Methods}

The experiment was performed at the base temperature ($\sim20$~mK) of a dilution refrigerator equipped with a 6-1-1~T vector magnet. Back-gate voltage was supplied by GS200 DC Voltage/Current source from Yokogawa. A standard low-frequency lock-in technique was used to measure the differential conductance $G=dI/dV|_{V_{bias}}=i_{ac}/v_{ac}$, with $V_{\mathrm{bias}}$ from Keithley 2450 and $v_{\mathrm{ac}}$ from NF LI5645 lock-in. The current $i_{ac}$ was amplified by an I/V converter SP983c from Basel Instruments and then measured by the lock-in. A small bias offset introduced by the I/V converter was numerically compensated for after the measurements. The linecut in Fig. \ref{fig:Fig1}\textbf{b} was produced by slightly smoothing the raw data using a 7-point Savitzky–Golay filter (3rd order).

\section{Data availability}

The data and code used in the generation of the main and supplementary figures are available on Zenodo with the identifier 10.5281/zenodo.19098398.

\section{Acknowledgments}

We thank Leo P. Kouwenhoven for helpful discussions. This work was funded by the Deutsche Forschungsgemeinschaft (DFG, German Research Foundation) under Germany’s Excellence Strategy -- Cluster of Excellence Matter and Light for Quantum Computing (ML4Q) EXC 2004/2 - 390534769 and by the DFG under CRC 1238 - 277146847 (subprojects A04 and B01).

\bibliography{bibliography}

@misc{schluck2024robust,
    title={Robust gap closing and reopening in topological-insulator {Josephson} junctions}, 
    author={Jakob Schluck and Ella Nikodem and Anton Montag and Alexander Ziesen and Mahasweta Bagchi and Fabian Hassler and Yoichi Ando},
    year={2024},
    eprint={2406.08265},
    archivePrefix={arXiv},
    primaryClass={cond-mat.supr-con}, 
}

@misc{nikodem2024topological,
    title={{Topological Insulator} nano-{SQUID}: {Flux}-tunable platform for topological superconductivity}, 
    author={Ella Nikodem and Jakob Schluck and Henry F. Legg and Max Geier and Michal Papaj and Mahasweta Bagchi and Liang Fu and Yoichi Ando},
    year={2025},
    eprint={2412.07993},
    archivePrefix={arXiv},
    primaryClass={cond-mat.mes-hall},
}

@article{hegde2020topological,
    title = {A topological {Josephson} junction platform for creating, manipulating, and braiding {Majorana} bound states},
    journal = {Ann. Phys.},
    volume = {423},
    pages = {168326},
    year = {2020},
    issn = {0003-4916},
    doi = {https://doi.org/10.1016/j.aop.2020.168326},
    url = {https://www.sciencedirect.com/science/article/pii/S0003491620302608},
    author = {Suraj S. Hegde and Guang Yue and Yuxuan Wang and Erik Huemiller and D.J. {Van Harlingen} and Smitha Vishveshwara}
}

@article{ben2015detecting,
    title = {Detecting {Majorana} modes in one-dimensional wires by charge sensing},
    author = {Ben-Shach, Gilad and Haim, Arbel and Appelbaum, Ian and Oreg, Yuval and Yacoby, Amir and Halperin, Bertrand I.},
    journal = {Phys. Rev. B},
    volume = {91},
    issue = {4},
    pages = {045403},
    numpages = {11},
    year = {2015},
    month = {Jan},
    publisher = {American Physical Society},
    doi = {10.1103/PhysRevB.91.045403},
    url = {https://link.aps.org/doi/10.1103/PhysRevB.91.045403}
}

@article{hu2007ge,
    title={A {Ge/Si} heterostructure nanowire-based double quantum dot with integrated charge sensor},
    author={Hu, Yongjie and Churchill, Hugh OH and Reilly, David J and Xiang, Jie and Lieber, Charles M and Marcus, Charles M},
    journal={Nat. Nanotechnol.},
    volume={2},
    number={10},
    pages={622--625},
    year={2007},
    publisher={Nature Publishing Group UK London},
    doi = {https://doi.org/10.1038/nnano.2007.302},
    url = {https://www.nature.com/articles/nnano.2007.302}
}

@article{mannila2019detecting,
    title = {Detecting parity effect in a superconducting device in the presence of parity switches},
    author = {Mannila, E. T. and Maisi, V. F. and Nguyen, H. Q. and Marcus, C. M. and Pekola, J. P.},
    journal = {Phys. Rev. B},
    volume = {100},
    issue = {2},
    pages = {020502},
    numpages = {5},
    year = {2019},
    month = {Jul},
    publisher = {American Physical Society},
    doi = {10.1103/PhysRevB.100.020502},
    url = {https://link.aps.org/doi/10.1103/PhysRevB.100.020502}
}

@article{vandriel2024charge,
  title = {Charge {Sensing} the {Parity} of an {Andreev Molecule}},
  author = {van Driel, David and Roovers, Bart and Zatelli, Francesco and Bordin, Alberto and Wang, Guanzhong and van Loo, Nick and Wolff, Jan Cornelis and Mazur, Grzegorz P. and Gazibegovic, Sasa and Badawy, Ghada and Bakkers, Erik P.A.M. and Kouwenhoven, Leo P. and Dvir, Tom},
  journal = {PRX Quantum},
  volume = {5},
  issue = {2},
  pages = {020301},
  numpages = {14},
  year = {2024},
  month = {Apr},
  publisher = {American Physical Society},
  doi = {10.1103/PRXQuantum.5.020301},
  url = {https://link.aps.org/doi/10.1103/PRXQuantum.5.020301}
}

@article{van2026single,
  title={Single-shot parity readout of a minimal {Kitaev} chain},
  author={van Loo, Nick and Zatelli, Francesco and Steffensen, Gorm O and Roovers, Bart and Wang, Guanzhong and Van Caekenberghe, Thomas and Bordin, Alberto and van Driel, David and Zhang, Yining and Huisman, Wietze D and others},
  journal={Nature},
  volume={650},
  number={8101},
  pages={334--339},
  year={2026},
  publisher={Nature Publishing Group UK London},
  doi = {doi.org/10.1038/s41586-025-09927-7},
  url = {https://www.nature.com/articles/s41586-025-09927-7}
}

@article{keller1996accuracy,
    author = {Keller, Mark W. and Martinis, John M. and Zimmerman, Neil M. and Steinbach, Andrew H.},
    title = {Accuracy of electron counting using a 7‐junction electron pump},
    journal = {Appl. Phys. Lett.},
    volume = {69},
    number = {12},
    pages = {1804-1806},
    year = {1996},
    month = {09},
    issn = {0003-6951},
    doi = {10.1063/1.117492},
    url = {https://doi.org/10.1063/1.117492}
}

@article{dresselhaus1994measurement,
    title = {Measurement of single electron lifetimes in a multijunction trap},
    author = {Dresselhaus, P. D. and Ji, L. and Han, Siyuan and Lukens, J. E. and Likharev, K. K.},
    journal = {Phys. Rev. Lett.},
    volume = {72},
    issue = {20},
    pages = {3226--3229},
    numpages = {0},
    year = {1994},
    month = {May},
    publisher = {American Physical Society},
    doi = {10.1103/PhysRevLett.72.3226},
    url = {https://link.aps.org/doi/10.1103/PhysRevLett.72.3226}
}

@article{aassime2001radio,
    author = {Aassime, A. and Gunnarsson, D. and Bladh, K. and Delsing, P. and Schoelkopf, R.},
    title = {Radio-frequency single-electron transistor: {Toward} the shot-noise limit},
    journal = {Appl. Phys. Lett.},
    volume = {79},
    number = {24},
    pages = {4031-4033},
    year = {2001},
    month = {12},
    issn = {0003-6951},
    doi = {10.1063/1.1424477},
    url = {https://doi.org/10.1063/1.1424477}
}

@article{schoelkopf1998radio,
    author = {R. J. Schoelkopf  and P. Wahlgren  and A. A. Kozhevnikov  and P. Delsing  and D. E. Prober },
    title = {{The Radio-Frequency Single-Electron Transistor (RF-SET): A Fast} and {Ultrasensitive Electrometer}},
    journal = {Science},
    volume = {280},
    number = {5367},
    pages = {1238-1242},
    year = {1998},
    doi = {10.1126/science.280.5367.1238},
    URL = {https://www.science.org/doi/abs/10.1126/science.280.5367.1238}
}

@article{katsnelson2006chiral,
    title={Chiral tunnelling and the {Klein} paradox in graphene},
    author={Katsnelson, Mikhail Iosifovich and Novoselov, Konstantin Sergejevic and Geim, Andre Konstantin},
    journal={Nat. Phys.},
    volume={2},
    number={9},
    pages={620--625},
    year={2006},
    publisher={Nature Publishing Group UK London},
    doi = {https://doi.org/10.1038/nphys384},
    url = {https://www.nature.com/articles/nphys384}
}

@article{zhang2010crossover,
  title={Crossover of the three-dimensional topological insulator {Bi2Se3} to the two-dimensional limit},
  author={Zhang, Yi and He, Ke and Chang, Cui-Zu and Song, Can-Li and Wang, Li-Li and Chen, Xi and Jia, Jin-Feng and Fang, Zhong and Dai, Xi and Shan, Wen-Yu and others},
  journal={Nat. Phys.},
  volume={6},
  number={8},
  pages={584--588},
  year={2010},
  publisher={Nature Publishing Group UK London},
  doi = {https://doi.org/10.1038/nphys1689},
  url = {https://www.nature.com/articles/nphys1689}
}

@article{munning2021quantum,
  title={Quantum confinement of the {Dirac} surface states in topological-insulator nanowires},
  author={M{\"u}nning, Felix and Breunig, Oliver and Legg, Henry F and Roitsch, Stefan and Fan, Dingxun and R{\"o}{\ss}ler, Matthias and Rosch, Achim and Ando, Yoichi},
  journal={Nat. Commun.},
  volume={12},
  number={1},
  pages={1038},
  year={2021},
  publisher={Nature Publishing Group UK London},
  doi = {https://doi.org/10.1038/s41467-021-21230-3},
  url = {https://www.nature.com/articles/s41467-021-21230-3}
}

@article{cho2012topological,
    author = {Cho, Sungjae and Kim, Dohun and Syers, Paul and Butch, Nicholas P. and Paglione, Johnpierre and Fuhrer, Michael S.},
    title = {{Topological Insulator Quantum Dot} with {Tunable Barriers}},
    journal = {Nano Lett.},
    volume = {12},
    number = {1},
    pages = {469-472},
    year = {2012},
    doi = {10.1021/nl203851g},
    URL = {https://doi.org/10.1021/nl203851g}
}

@article{jing2019single,
    author = {Jing, Yumei and Huang, Shaoyun and Wu, Jinxiong and Meng, Mengmeng and Li, Xiaobo and Zhou, Yu and Peng, Hailin and Xu, Hongqi},
    title = {A {Single-Electron Transistor Made} of a {3D Topological Insulator Nanoplate}},
    journal = {Adv. Mater.},
    volume = {31},
    number = {42},
    pages = {1903686},
    doi = {https://doi.org/10.1002/adma.201903686},
    url = {https://advanced.onlinelibrary.wiley.com/doi/abs/10.1002/adma.201903686},
    year = {2019}
}

@article{arakane2012tunable,
    title={Tunable {Dirac} cone in the topological insulator {Bi2-xSbxTe3-ySey}},
    author={Arakane, T and Sato, T and Souma, S and Kosaka, K and Nakayama, K and Komatsu, M and Takahashi, T and Ren, Zhi and Segawa, Kouji and Ando, Yoichi},
    journal={Nat. Commun.},
    volume={3},
    number={1},
    pages={636},
    year={2012},
    publisher={Nature Publishing Group UK London},
    doi = {https://doi.org/10.1038/ncomms1639},
    url = {https://www.nature.com/articles/ncomms1639}
}

@article{cook2011majorana,
  title = {Majorana fermions in a topological-insulator nanowire proximity-coupled to an $s$-wave superconductor},
  author = {Cook, A. and Franz, M.},
  journal = {Phys. Rev. B},
  volume = {84},
  issue = {20},
  pages = {201105},
  numpages = {4},
  year = {2011},
  month = {Nov},
  publisher = {American Physical Society},
  doi = {10.1103/PhysRevB.84.201105},
  url = {https://link.aps.org/doi/10.1103/PhysRevB.84.201105}
}

@article{razmadze2019radio,
    title = {{Radio-Frequency Methods} for {Majorana-Based Quantum Devices: Fast Charge Sensing and Phase-Diagram Mapping}},
    author = {Razmadze, Davydas and Sabonis, Deividas and Malinowski, Filip K. and M\'enard, Gerbold C. and Pauka, Sebastian and Nguyen, Hung and van Zanten, David M.T. and O\ensuremath{'}Farrell, Eoin C.T. and Suter, Judith and Krogstrup, Peter and Kuemmeth, Ferdinand and Marcus, Charles M.},
    journal = {Phys. Rev. Appl.},
    volume = {11},
    issue = {6},
    pages = {064011},
    numpages = {9},
    year = {2019},
    month = {Jun},
    publisher = {American Physical Society},
    doi = {10.1103/PhysRevApplied.11.064011},
    url = {https://link.aps.org/doi/10.1103/PhysRevApplied.11.064011}
}

@article{feng2025long,
    title={Long-range crossed {Andreev} reflection in a topological insulator nanowire proximitized by a superconductor},
    author={Feng, Junya and Legg, Henry F and Bagchi, Mahasweta and Loss, Daniel and Klinovaja, Jelena and Ando, Yoichi},
    journal={Nat. Phys.},
    pages={1--8},
    year={2025},
    publisher={Nature Publishing Group UK London},
    doi = {https://doi.org/10.1038/s41567-025-02806-y},
    url = {https://www.nature.com/articles/s41567-025-02806-y}
}

@article{wunnicke2006gate,
    author = {Wunnicke, Olaf},
    title = {Gate capacitance of back-gated nanowire field-effect transistors},
    journal = {Appl. Phys. Lett.},
    volume = {89},
    number = {8},
    pages = {083102},
    year = {2006},
    month = {08},
    issn = {0003-6951},
    doi = {10.1063/1.2337853},
    url = {https://doi.org/10.1063/1.2337853}
}

@article{hofheinz2006individual,
    title={Individual charge traps in silicon nanowires: {Measurements} of location, spin and occupation number by {Coulomb} blockade spectroscopy},
    author={Hofheinz, M and Jehl, X and Sanquer, M and Molas, G and Vinet, M and Deleonibus, S},
    journal={Eur. Phys. J. B},
    volume={54},
    number={3},
    pages={299--307},
    year={2006},
    publisher={Springer},
    doi = {https://doi.org/10.1140/epjb/e2006-00452-x},
    url = {https://link.springer.com/article/10.1140/epjb/e2006-00452-x}
}

@article{grupp2001dynamical,
    title = {{Dynamical Offset Charges} in {Single-Electron Transistors}},
    author = {Grupp, D. E. and Zhang, T. and Dolan, G. J. and Wingreen, Ned S.},
    journal = {Phys. Rev. Lett.},
    volume = {87},
    issue = {18},
    pages = {186805},
    numpages = {4},
    year = {2001},
    month = {Oct},
    publisher = {American Physical Society},
    doi = {10.1103/PhysRevLett.87.186805},
    url = {https://link.aps.org/doi/10.1103/PhysRevLett.87.186805}
}

@article{villis2014direct,
    author = {Villis, B. J. and Orlov, A. O. and Barraud, S. and Vinet, M. and Sanquer, M. and Fay, P. and Snider, G. and Jehl, X.},
    title = {Direct detection of a transport-blocking trap in a nanoscaled silicon single-electron transistor by radio-frequency reflectometry},
    journal = {Appl. Phys. Lett.},
    volume = {104},
    number = {23},
    pages = {233503},
    year = {2014},
    month = {06},
    issn = {0003-6951},
    doi = {10.1063/1.4883228},
    url = {https://doi.org/10.1063/1.4883228}
}

@article{knispel2017charge,
    title = {Charge puddles in the bulk and on the surface of the topological insulator {${\mathrm{BiSbTeSe}}_{2}$} studied by scanning tunneling microscopy and optical spectroscopy},
    author = {Knispel, T. and Jolie, W. and Borgwardt, N. and Lux, J. and Wang, Zhiwei and Ando, Yoichi and Rosch, A. and Michely, T. and Gr\"uninger, M.},
    journal = {Phys. Rev. B},
    volume = {96},
    issue = {19},
    pages = {195135},
    numpages = {7},
    year = {2017},
    month = {Nov},
    publisher = {American Physical Society},
    doi = {10.1103/PhysRevB.96.195135},
    url = {https://link.aps.org/doi/10.1103/PhysRevB.96.195135}
}

@article{brede2024characterizing,
    title = {Characterizing the chemical potential disorder in the topological insulator ${({\mathrm{Bi}}_{1\ensuremath{-}x}{\mathrm{Sb}}_{x})}_{2}{{\mathrm{Te}}}_{3}$ thin films},
    author = {Brede, Jens and Bagchi, Mahasweta and Greichgauer, Adrian and Uday, Anjana and Bliesener, Andrea and Lippertz, Gertjan and Yazdanpanah, Roozbeh and Taskin, Alexey and Ando, Yoichi},
    journal = {Phys. Rev. Mater.},
    volume = {8},
    issue = {10},
    pages = {104202},
    numpages = {12},
    year = {2024},
    month = {Oct},
    publisher = {American Physical Society},
    doi = {10.1103/PhysRevMaterials.8.104202},
    url = {https://link.aps.org/doi/10.1103/PhysRevMaterials.8.104202}
}

@article{guttinger2009electron,
    title = {{Electron-Hole Crossover} in {Graphene Quantum Dots}},
    author = {G\"uttinger, J. and Stampfer, C. and Libisch, F. and Frey, T. and Burgd\"orfer, J. and Ihn, T. and Ensslin, K.},
    journal = {Phys. Rev. Lett.},
    volume = {103},
    issue = {4},
    pages = {046810},
    numpages = {4},
    year = {2009},
    month = {Jul},
    publisher = {American Physical Society},
    doi = {10.1103/PhysRevLett.103.046810},
    url = {https://link.aps.org/doi/10.1103/PhysRevLett.103.046810}
}

@article{zimmerman1997modulation,
    title = {Modulation of the charge of a single-electron transistor by distant defects},
    author = {Zimmerman, Neil M. and Cobb, Jonathan L. and Clark, Alan F.},
    journal = {Phys. Rev. B},
    volume = {56},
    issue = {12},
    pages = {7675--7678},
    numpages = {0},
    year = {1997},
    month = {Sep},
    publisher = {American Physical Society},
    doi = {10.1103/PhysRevB.56.7675},
    url = {https://link.aps.org/doi/10.1103/PhysRevB.56.7675}
}

@article{borgwardt2016self,
    title = {Self-organized charge puddles in a three-dimensional topological material},
    author = {Borgwardt, N. and Lux, J. and Vergara, I. and Wang, Zhiwei and Taskin, A. A. and Segawa, Kouji and van Loosdrecht, P. H. M. and Ando, Yoichi and Rosch, A. and Gr\"uninger, M.},
    journal = {Phys. Rev. B},
    volume = {93},
    issue = {24},
    pages = {245149},
    numpages = {12},
    year = {2016},
    month = {Jun},
    publisher = {American Physical Society},
    doi = {10.1103/PhysRevB.93.245149},
    url = {https://link.aps.org/doi/10.1103/PhysRevB.93.245149}
}

@phdthesis{hofheinz2006coulomb,
  TITLE = {Coulomb blockade in silicon nanowire MOSFETs},
  AUTHOR = {Hofheinz, Max},
  URL = {https://theses.hal.science/tel-00131052},
  SCHOOL = {{Universit{\'e} Joseph-Fourier - Grenoble I}},
  YEAR = {2006},
  MONTH = Dec,
  TYPE = {Theses},
  HAL_ID = {tel-00131052},
  HAL_VERSION = {v1},
}

@article{beenakker1991theory,
  title = {Theory of {Coulomb}-blockade oscillations in the conductance of a quantum dot},
  author = {Beenakker, C. W. J.},
  journal = {Phys. Rev. B},
  volume = {44},
  issue = {4},
  pages = {1646--1656},
  numpages = {0},
  year = {1991},
  month = {Jul},
  publisher = {American Physical Society},
  doi = {10.1103/PhysRevB.44.1646},
  url = {https://link.aps.org/doi/10.1103/PhysRevB.44.1646}
}

\end{document}


\title{Supplementary Information: Topological insulator single-electron transistors for charge sensing applications}

\author{Omargeldi Atanov}
\thanks{These authors contributed equally to this work. \newline Correspondence: \href{mailto:atanov@ph2.uni-koeln.de}{atanov@ph2.uni-koeln.de}}
\affiliation{Physics Institute II, University of Cologne, D-50937 K{\"o}ln, Germany}

\author{Junya Feng}
\thanks{These authors contributed equally to this work. \newline Correspondence: \href{mailto:atanov@ph2.uni-koeln.de}{atanov@ph2.uni-koeln.de}}
\affiliation{Physics Institute II, University of Cologne, D-50937 K{\"o}ln, Germany}

\author{Jens Brede}
\affiliation{Physics Institute II, University of Cologne, D-50937 K{\"o}ln, Germany}

\author{Oliver Breunig}
\affiliation{Physics Institute II, University of Cologne, D-50937 K{\"o}ln, Germany}

\author{Yoichi Ando}
\affiliation{Physics Institute II, University of Cologne, D-50937 K{\"o}ln, Germany}

\maketitle

\section{Fast Fourier transform analysis of the Coulomb resonances}

\begin{figure*}[htbp]
    \centering
    \includegraphics[width=\textwidth, trim={0cm 0cm 0cm 0cm},clip]{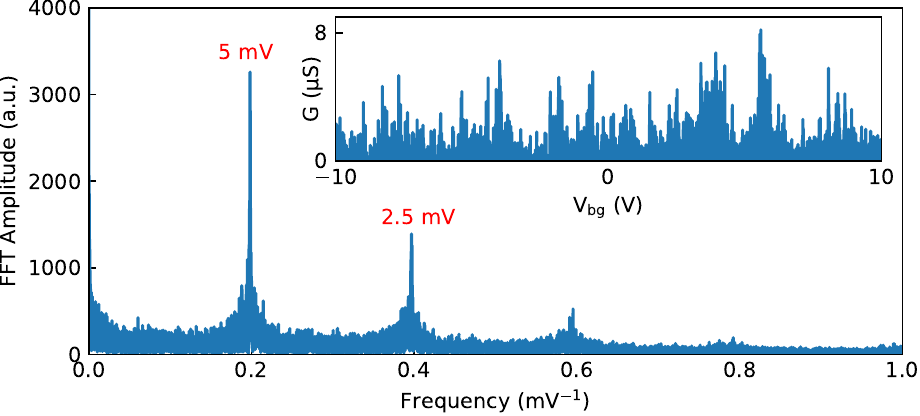}
    \caption{FFT analysis of the conductance oscillations from Device 1 presented in the main text. A single dominant frequency corresponding to $\delta V_{\mathrm{bg}}\approx 5$~mV periodicity is clearly identified across $20$~V back-gate voltage range. Inset shows the zero-bias Coulomb resonance peaks across the full range of the measurement.
    }
    \label{fig:FigS1}
\end{figure*}

We perform a fast Fourier transform (FFT) analysis of the conductance oscillations in Device 1 over a wide back-gate voltage range and identify a single dominant frequency in the spectrum as shown in Fig. \ref{fig:FigS1}. This indicates that the Coulomb charging energy is the only relevant energy scale and that the device operates as a metallic SET.

\newpage

\section{More devices showing Coulomb blockade}

\begin{figure*}[htbp]
    \centering
    \includegraphics[width=0.55\textwidth, trim={0cm 0cm 0cm 0cm},clip]{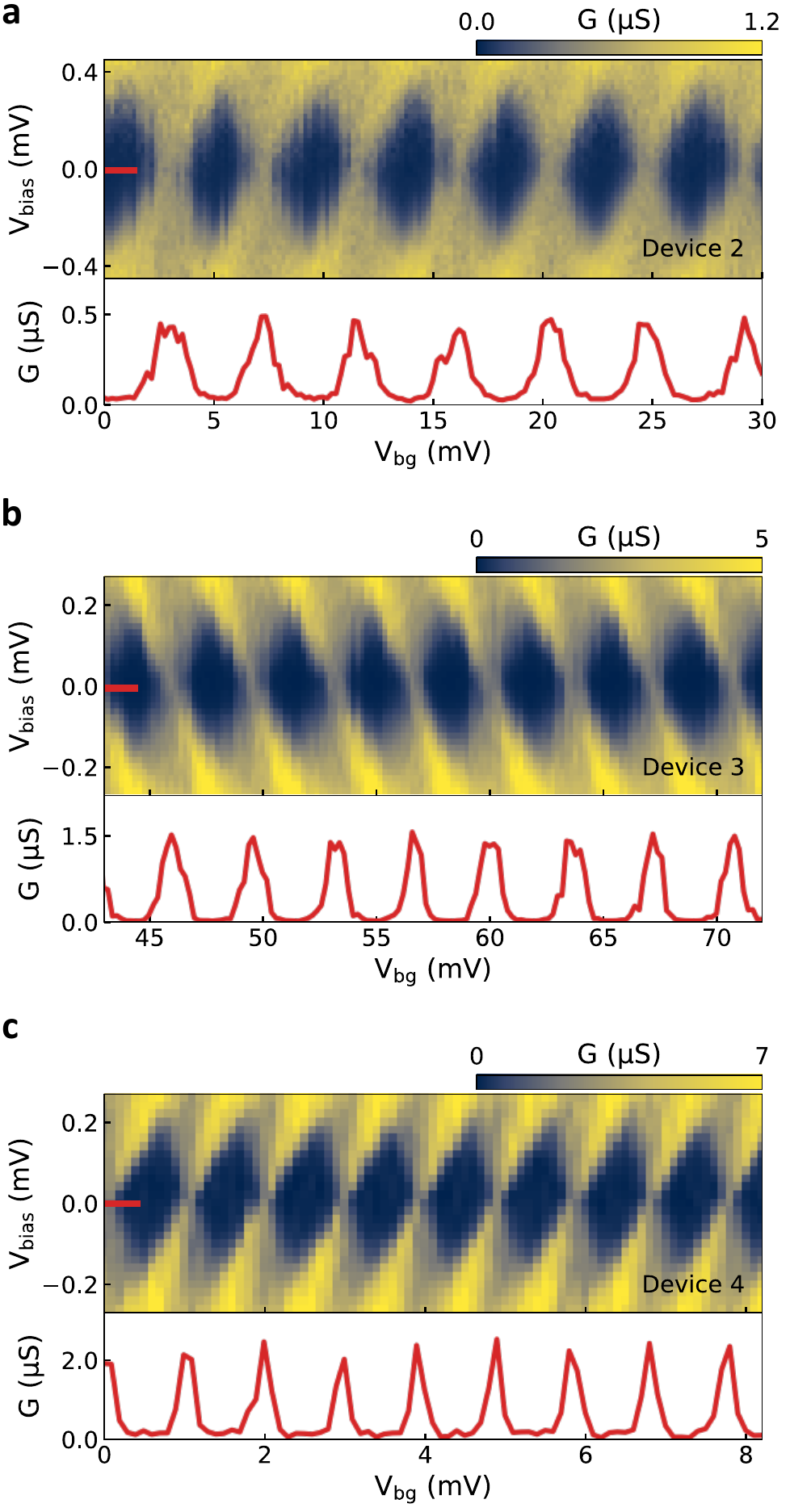}
    \caption{Charge stability diagrams of \textbf{a} Device 2, \textbf{b} Device 3, and \textbf{c} Device 4 all of which exhibit clear Coulomb diamonds. The lower panel in each plot shows the linecut taken at zero-bias voltage showing the Coulomb resonances.}
    \label{fig:FigS2}
\end{figure*}

In Fig. \ref{fig:FigS2}, we present the stability diagrams of 3 more devices (Device 2, Device 3, and Device 4) that show Coulomb-blockaded transport. All devices have been fabricated on the same chip as the Device 1 discussed in the main text.

\newpage

\section{Resonance shifts in Device 3}

\begin{figure*}[htbp]
    \centering
    \includegraphics[width=\textwidth, trim={0cm 0cm 0cm 0cm},clip]{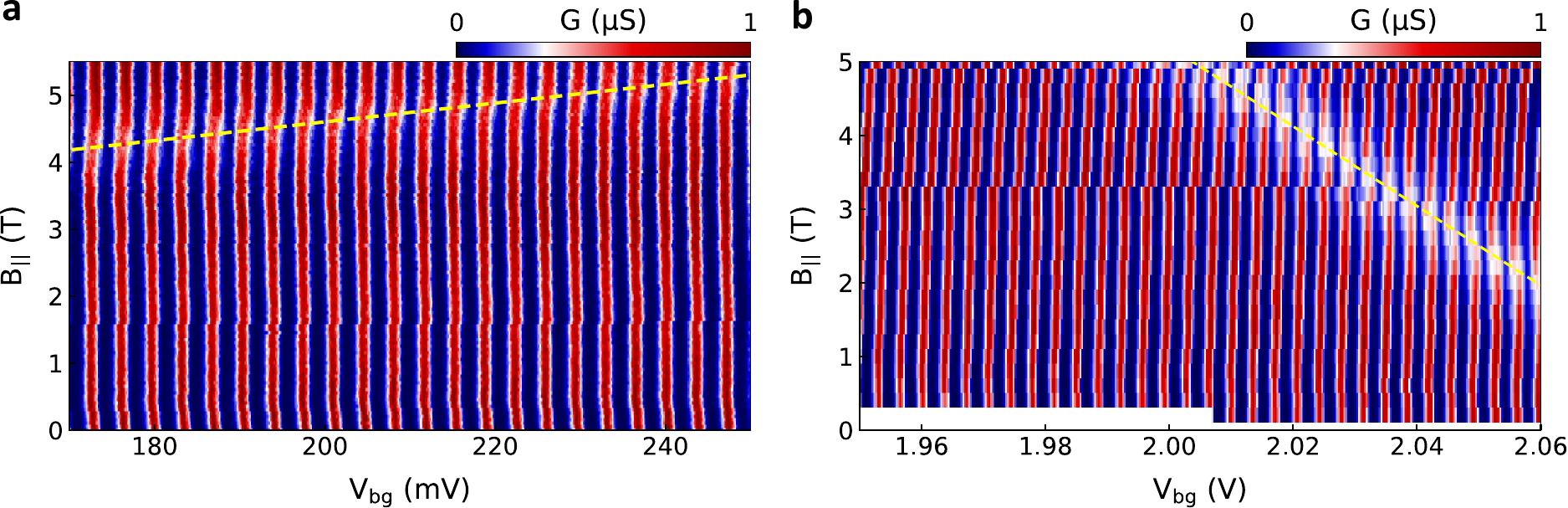}
    \caption{Second device showing trap-induced resonance shifts in two different regions. \textbf{a} A `spin-up' trap slightly shifts the resonances of the main SET island which is accompanied by the reduction of the conductance peaks. \textbf{b} A `spin-down' trap is identified at a higher gate voltage with stronger coupling to the gate electrode.}
    \label{fig:FigS3}
\end{figure*}

In Fig. \ref{fig:FigS3}, we present the zero-bias conductance of Device 3 in the ($V_{\mathrm{bg}}$, $B_{||}$) plane. Although quite subtle, we are able to resolve two regions where there are systematic shifts of the Coulomb resonances, recognized mainly by the suppression of the conductance peaks. Consistent with the data in the main text, the trap signature shifts linearly with the applied $B_{||}$ field. A diverging color scale is used to highlight the faint trap signatures that are difficult to identify in a linear color scale. A few abrupt resonance jumps observed in the plots are due to the charge rearrangements in the electrostatic environment.

\newpage

\section{Local side-gate operation}

\begin{figure*}[htbp]
    \centering
    \includegraphics[width=\textwidth, trim={0cm 0cm 0cm 0cm},clip]{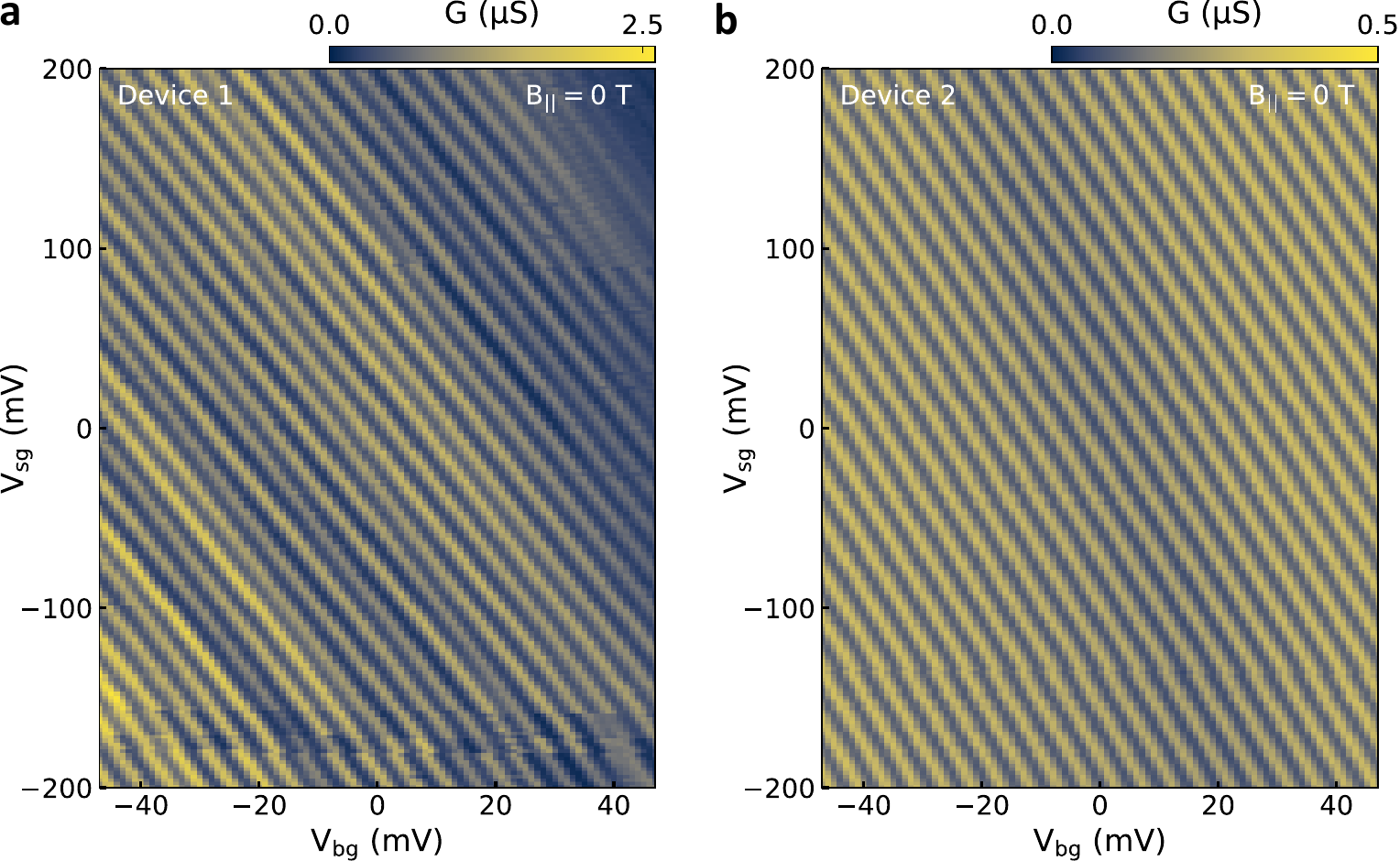}
    \caption{Demonstration of the side-gate operation for \textbf{a} Device 1 and \textbf{b} Device 2. In both devices, the side-gate modulates the chemical potential of the TINW island with a smaller lever arm compared to the back-gate.}
    \label{fig:FigS4}
\end{figure*}

In Fig. \ref{fig:FigS4}, we demonstrate our capability of tuning the chemical potential of the TINW islands via a local side-gate. The figure shows the differential conductance $G$ as a function of back-gate $V_{\mathrm{bg}}$ and side-gate $V_{\mathrm{sg}}$ voltages measured at zero $B_{||}$ field. The resonances run diagonally across the ($V_{\mathrm{bg}}$, $V_{\mathrm{sg}}$) plane, confirming that both gates are tuning the chemical potential simultaneously.

\newpage

\section{Details of the numerical simulation}

\subsection*{Electrostatic Model of the SET-trap System}

We model the device as a SET island capacitively coupled to a localized charge trap. The electrostatic energy of the coupled system is written as \cite{hofheinz2006coulomb}:
\begin{equation}
U(n, n_t)=E_C(N-n+\beta_t(N_t-n_t))^2 + E_t(N_t-n_t)^2
\end{equation}
where $n$ is the integer charge on the SET island and $n_t$ denotes the trap state occupation. The SET charging energy is

\begin{equation}
E_C=\frac{e^2}{2C_\Sigma},
\qquad C_\Sigma=C_s+C_d+C_g ,
\end{equation}

and the effective offset charge is

\begin{equation}
N = \frac{C_s V_s + C_d V_d + C_g V_g + q_0}{e},
\end{equation}

where $C_s$, $C_d$, and $C_g$ are the source, drain, and gate capacitances of the SET and $q_0$ is the background charge that is tuned to shift the Coulomb resonances in gate-voltage axis. The trap charging energy is

\begin{equation}
E_t=\frac{e^2}{2(C_{ts}+C_{tg}+C_c)},
\end{equation}

with $C_{ts}$ and $C_{tg}$ the trap--source and trap--gate capacitances, and $C_c$ the mutual capacitance between the trap and the SET island. The dimensionless trap offset charge is

\begin{equation}
N_t=\frac{C_{ts}V_s+C_{tg}(V_g+K)}{e}.
\end{equation}

where $K$ is a constant gate-voltage offset to compensate for the internal energy difference between filled and empty trap. The coupling parameter

\begin{equation}
\beta_t=\frac{C_c}{C_{ts}+C_{tg}+C_c}
\end{equation}

quantifies the induced charge on the SET island due to trap occupation and is directly related to the experimentally observed shift of Coulomb resonances,
$\beta_t=\Delta V_g/\delta V_g$. This form of the electrostatic energy follows standard SET--trap models and corresponds to a capacitive coupling treatment in which higher-order cross terms from the full capacitance matrix are absorbed into the definition of $\beta_t$.

\subsection*{Spinful Trap and Magnetic-Field Dependence}

The trap is assumed to host a single spinful electronic state. In a magnetic field $B$, the trap energy is shifted by the Zeeman energy

\begin{equation}
E_Z = \pm \frac{1}{2} g \mu_B B,
\end{equation}

where $g\simeq 2$ is the effective $g$-factor of the nanowire used in the experiment. Since tunneling of an electron changes the trap spin by $\Delta S_z=\pm 1/2$, the spin degree of freedom can be absorbed into the sign of $E_Z$, and an explicit spin index is omitted. The Zeeman shift of the trap state is modeled as an effective shift of the trap gate offset,

\begin{equation}
N_t=\frac{C_{ts}V_s + C_{tg}\left(V_g+K-\dfrac{E_Z}{e\alpha_t}\right)}{e},
\end{equation}

where

\begin{equation}
\alpha_t=\frac{C_{tg}}{C_{ts}+C_{tg}+C_c}
\end{equation}

is the trap gate lever arm. This expression assumes that the magnetic field predominantly affects the trap state energy and that its effect can be mapped onto an equivalent gate-voltage shift via $\alpha_t$.

\subsection*{Charge States and Transport Formalism}

Charge states are labeled as $(n,n_t)$, corresponding to the SET island charge and the trap state occupation. We simulate transport through 22 different charge states by limiting the number of electrons on the SET island to 10 and on the trap state to 1, i.e. $n\in\{0,1, ...,10\}$ and $n_t\in\{0,1\}$. Transport is described using a Pauli master equation for the occupation probabilities $P_{n,n_t}$ \cite{beenakker1991theory},

\begin{equation}
\frac{\partial \bm{P}}{\partial t}=\Gamma \bm{P},
\end{equation}

where $\Gamma$ is the transition-rate matrix. In the stationary limit,

\begin{equation}
\Gamma \bm{P}=0, \qquad
\sum_{n,n_t} P_{n,n_t}=1 .
\end{equation}

The current through the leads is calculated as:

\begin{equation}
I_{s,d} = - e\sum_{n,n_t}\sum_{n',n_t'}\left(N_{n',n_t'}-N_{n,n_t}\right)
\Gamma^{(s,d)}_{(n',n_t')\leftarrow(n,n_t)}P_{n,n_t},
\end{equation}

where $N_{n,n_t}$ is the total number of electrons in a given $(n,n_t)$ charge state. The differential conductance $G$ is given by the numerical derivative of this current with respect to the bias voltage. Since the trap state is a single discrete level, transition rates in or out of it are modeled using the Fermi function,

\begin{equation}
\Gamma^{(s,d)}_{(n',n_t')\leftarrow(n,n_t)}=\gamma_t
f\!\left(\frac{U_{n',n_t'}-U_{n,n_t}+\mu_{s,d}}{k_B T}\right),
\end{equation}

with

\begin{equation}
f(x)=\frac{1}{e^x+1}.
\end{equation}

Here $\gamma_t$ is the constant transmission coefficient of the trap towards the leads or the SET island, $T$ is the electron temperature and $\mu_{s,d}=-eV_{s,d}$ is the chemical potential of the source and drain leads with $V_{s}-V_{d}=V_{\mathrm{bias}}$.
Transition rates between the SET island and the reservoirs, both of which have continuous energy spectrum, are modeled using a Fermi auto-convolution form,

\begin{equation}
\Gamma^{(s,d)}_{(n',n_t')\leftarrow(n,n_t)}=\gamma_t
f^*\!\left(\frac{U_{n',n_t'}-U_{n,n_t}+\mu_{s,d}}{k_B T}\right),
\end{equation}

with

\begin{equation}
f^*(x)=(f*f)(x)=\frac{x}{e^x-1}.
\end{equation}

Here $\gamma_l^{s,d}$ is the constant transmission coefficient of the SET island towards source and drain leads. Forward and backward tunneling processes are treated separately through the sign of the energy difference $U_{n',n_t'}-U_{n,n_t}$.

\subsection*{Parameter Extraction}

From the experimentally observed shift of Coulomb peaks in region `1', $\beta_t\simeq 0.5$, one obtains

\begin{equation}
C_c \simeq C_{ts}+C_{tg}.
\end{equation}

Using this relation together with the definition of $\alpha_t$,

\begin{equation}
\alpha_t=\frac{C_{tg}}{2(C_{ts}+C_{tg})},
\end{equation}

the trap gate capacitance follows as

\begin{equation}
C_{tg}=\frac{2\alpha_t C_{ts}}{1-2\alpha_t},
\end{equation}

and the mutual capacitance as

\begin{equation}
C_c=C_{ts}\frac{1+2\alpha_t}{1-2\alpha_t}.
\end{equation}

The measured slope of conductance resonances in the $(V_\mathrm{bg},B_{||})$ plane provides an independent determination of the trap lever arm,

\begin{equation}
\alpha_t=\pm \frac{1}{2}\frac{g\mu_B}{e}
\frac{\partial B_{||}}{\partial V_\mathrm{bg}}.
\end{equation}

\subsection*{Simulation Parameters}

\begin{table}[h]
\caption{The source and the values of the parameters used in the simulations.}
\centering
\begin{tabular}{lll}
\toprule
Parameter & Value & Source \\
\midrule
$E_C$ & $0.34~\mathrm{meV}$ & Extracted from measured Coulomb diamond height at $\sim 5$~T. \\
$C_g$ & $32~\mathrm{aF}$ & Extracted from Coulomb peak spacings. \\
$C_s$ & $280~\mathrm{aF}$ & Extracted from $C_\Sigma$ determined by $E_C$. \\
$C_d$ & $160~\mathrm{aF}$ & Extracted from $C_\Sigma$ determined by $E_C$. \\
$\alpha_t$ & $2.15 \times 10^{-3}$ & Extracted from the slope of the Zeeman shift in $G(V_\mathrm{bg}, B_{||})$. \\
$\beta_t$ & $0.5$ & Extracted from observed resonance shifts of trap `2'. \\
$C_{ts}$ & $100~\mathrm{aF}$ & Independently set to roughly match the trap-induced distortions. \\
$C_c$ & $100.43~\mathrm{aF}$ & Calculated using $\alpha_t$, $\beta_t$, and $C_{ts}$ values. \\
$C_{tg}$ & $0.43~\mathrm{aF}$ & Calculated using $\alpha_t$, $\beta_t$, and $C_{ts}$ values. \\
$T$ & $120~\mathrm{mK}$ & Independently set to roughly match the zero-bias conductance. \\
$\gamma^{s,d}_l$ & $6 \times 10^8$ & Independently set to roughly match high-bias conductance. \\
$\gamma_t$ & $5 \times 10^5$ & Independently set for a small trap transition rate. \\
\bottomrule
\end{tabular}
\end{table}

\bibliography{bibliography}